\documentstyle[preprint,prb,aps]{revtex}
\tightenlines
\begin{document}
\title{Nucleation in dilute $^3$He-$^4$He liquid
mixtures at low temperatures}
\author{M. Barranco$^1$\cite{perm}, M. Guilleumas$^1$, D. M. Jezek$^2$,
R. J. Lombard$^3$, J. Navarro$^4$, and M. Pi$^5$}

\address{$^1$Dipartimento di Fisica, Universit\`a di Trento, and
Istituto Nazionale Fisica della Materia.
38050 Povo, Italy}
\address{$^2$Departamento de F\'{\i}sica,
Universidad de Buenos Aires. 1428 Buenos Aires, Argentine}
\address{$^3$Institut de Physique Nucl\'eaire, Division de Physique
Th\'eorique. 91406 Orsay, France}
\address{$^4$Institut de F\'{\i}sica Corpuscular, CSIC. 46100 Burjassot,
Spain}
\address{$^5$Departament ECM, Facultat de F\'{\i}sica,
Universitat de Barcelona. 08028 Barcelona, Spain}
\date{\today}

\maketitle

\begin{abstract}

We present a study of phase separation from supersaturated $^3$He-$^4$He
liquid mixtures at low temperatures addressing both the degree of
critical supersaturation $\Delta x_{cr}$ and the thermal-to-quantum
crossover temperature $T^*$ for the nucleation process. Two different
nucleation seeds are investigated, namely $^3$He droplets and $^4$He
vortex lines with cores filled with $^3$He.
We have found that the experimental
$T^*$ is reproduced when we consider that
 nucleation proceeds from $^3$He droplets, whereas
$\Delta x_{cr}$ is reproduced when we consider
 $^4$He vortex lines filled with $^3$He.
However, neither nucleation
configuration is able to simultaneously reproduce the current
experimental information on $\Delta x_{cr}$ and $T^*$.

\end{abstract}


\section{Introduction}

Supersaturated $^3$He-$^4$He liquid mixtures have been recognized as an
ideal system to study nucleation phenomena. On the one hand, these
mixtures can be made extremely pure, and on the other hand, $^4$He
covers the walls of the experimental cells, preventing  undesired
heterogeneous nucleation that is difficult to avoid in other
substances. Moreover, the fact that  helium remains liquid up to zero
temperature ($T$) has raised expectations that  the
transition from thermal to quantum nucleation could be observed
as $T$ decreases.

The first experiments on metastable helium mixtures were carried out in
the sixties \cite{Lan69,Wat69}, but one had to wait more than twenty
years for a systematic study of this phenomenon
\cite{Sat91,Mik91,Sat92,Mai92,Mik94,Sat94,Cha98}. Meanwhile, the
kinetics of nucleation in dilute helium mixtures was theoretically
addressed by Lifshitz et al \cite{Lif78},
 much along the line
of the classical work of Lifshitz and Kagan \cite{Lif72}.

At very low temperature and at zero pressure, the concentration of $^3$He
in the mixture at saturation is\cite{Lan69} $x_s \sim 6.6\%$. However,
supersaturated $^3$He-$^4$He mixtures can be found in a metastable state
for concentrations $x$  above the saturation value. A first estimate of
the degree of supersaturation $\Delta x_{cr} \equiv x - x_s$ resulted
from extrapolation of the measured $^3$He chemical potential excess
$\Delta \mu_3$ along the demixing line. This extrapolation yielded
$\partial \Delta \mu_3 / \partial x \geq 0$ up to  values of
the $^3$He concentration
$x \gtrsim $ 16 \%, thus giving \cite{Sel69} $\Delta x_{cr} \sim 10\%$.
Large values of critical supersaturation have also been
obtained in microscopic \cite{Kro93} and density functional
\cite{Gui95} calculations.

An intriguing observation is the small degree of
critical supersaturation attained
in recent experiments on phase separation
in supersaturated helium mixtures. It was found to be of the order of
1 \% in the experiments carried out by the Ukrainian group
\cite{Mik91,Mai92,Mik94,Cha98}, and below 0.5 \% in those
of the Japanese group \cite{Sat91,Sat92,Sat94}.
As a matter of fact, the  measurements in the sixties\cite{Lan69,Wat69}
 were also yielding  small $\Delta x_{cr}$.

A possible explanation of the discrepancy between theory and experiment
is that phase separation at small supersaturations
may be triggered by heterogeneous
nucleation on vortex lines. Indeed,  at the low temperatures
involved in the experiments $^4$He is still superfluid due to
 the limited solubility of $^3$He into $^4$He. This mechanism
was proposed in Ref. \onlinecite{Jez95}, and it has been further
elaborated\cite{Jez97,Jez98}.
It turns out that the presence of vortex lines
in the mixture decreases the degree of critical supersaturation from a
value $\Delta x_{cr} \sim 10 \%$ to $\sim$ 1 \%.
At present, both experimental groups seem to
have adhere to this explanation of their results \cite{Bur96,Bur98,Cha98b}.

Besides critical supersaturation, another experimental  quantity
of most interest is the temperature at which the
nucleation process changes from being a thermal one to a quantal one.
This crossover
temperature $T^*$ has been determined  by measuring the
temperature below which the degree of critical supersaturation becomes
almost $T$ independent\cite{Sat92};
 it is of the order of 10 mK. Obviously, the
theoretical description of the process cannot be considered as
achieved until
$\Delta x_{cr}$ and $T^*$ are satisfactorily obtained within the same
framework.

The present aim is to complete our previous works with
the calculation of the crossover temperature using the same basic
ingredients that we have used to obtain the critical supersaturation.
To this end, we have resorted to
a dynamical method, namely the functional-integral
approach (FIA), that  proved\cite{Gui96} to be well suited to
quantitatively describe quantum cavitation in liquid $^4$He.

\section{Formalism}

The segregation of $^3$He from a dilute $^3$He-$^4$He liquid mixture at
low $T$ is a first order phase transition. This means that in
order to nucleate the pure $^3$He phase, the system has to overcome an
energy barrier $\Delta\Omega$. Thus, the mixture may be at $^3$He
concentrations  above the one corresponding to the two-phase
equilibrium value at the given pressure ($P$) and temperature
\cite{Ebn70}.

At high enough $T$, segregation is thermally activated and
the energy barrier is overcome by the energy provided
to the system by a thermal bath. In this case, the nucleation rate,
i.e., the number of critical nucleation seeds formed per unit volume and
time is given by
\begin{equation}
J_T = J_{0T}\, exp \left[-\frac{\Delta\Omega_{max}}{k_B T}\right] \; ,
\label{eq1}
\end{equation}
where the prefactor $J_{0T}$ depends on the dynamics of
the nucleation process, $k_B$ is the Boltzmann's constant and
$\Delta\Omega_{max}$ is the height of the energy barrier. At low
$T$, segregation by thermal activation is no longer possible.
However, it can proceed by quantum tunneling: the metastable liquid mixture
`tunnels' through the energy barrier.
In the limit of zero temperature, the transition is purely quantal, but
below $T^*$ thermally assisted tunneling also occurs.
This is one of the many
thermally assisted quantum tunneling processes occuring in physics
(see for example Ref. \onlinecite{Chu92} and refs. therein).
For $T < T^*$ the tunneling rate is
\begin{equation}
J_Q = J_{0Q} \,exp(-S^Q)     \;  ,
\label{eq2}
\end{equation}
where $exp(-S^Q)$ is the tunneling probability, and the prefactor
$J_{0Q}$ is of the order of the number of nucleation sites per unit
volume times an attempting frequency.

To obtain the tunneling probability one formally starts from the
statistical
average of the transition probability over a time $t = t_f - t_i$:
\begin{equation}
P = \frac{\sum_{i,f}|\langle \Psi_f | exp \left[ - \frac{\imath}
{\hbar} \int^{t_f}_{t_i} d t \hat{\cal H} \right] |
\Psi_i \rangle |^2 exp (-E_i/k_B T)}
{\sum_i exp(-E_i/k_B T)}  \; ,
\label{eq3}
\end{equation}
where $\hat{\cal H}$ is the Hamiltonian of the system, $E_i$ are the
energy
states, and $\Psi_i$ and $\Psi_f$ are the wave functions of the initial
and final states. This expression can be written in a more workable form
by using the path integral formulation of quantum mechanics, and its
connection with statistical mechanics. According to Feynman \cite{Fey72}
(see also Refs. \onlinecite{Lan67,Col77,Cal77}), $P$ can be written as
the functional integral
\begin{equation}
 \int  D [q(\tau)] \,exp \left[-\frac{1}{\hbar} \oint d \tau
{\cal L}[q(\tau)] \right] \; ,
\label{eq4}
\end{equation}
where ${\cal L}[q(\tau)]$ is the imaginary-time ($\tau = \imath t$)
classical Lagrangian of the system
and $D[q(\tau)]$ denotes integration over all periodic trajectories
$q(\tau)$ with period $\tau_p = \hbar / k_B T$
in the potential well that results from inverting the energy  barrier.
 The integral in the exponent of Eq. (\ref{eq4})
is the imaginary-time action $S(T) = \oint d \tau {\cal L}$ evaluated
over the period $\tau_p$. In the semiclassical limit
$S(T) >> \hbar$, the trajectory that contributes the most from
all possible periodic orbits is the one which minimizes the action.
It leads to Eq. (\ref{eq2}) with  $S^Q  = S_{min}(T)/ \hbar$, where
$S_{min}$ is the minimum action.

The practical usefulness of Eq. (\ref{eq4}) for the problem at
hand is that physical insight allows one to guess  $\cal L$ as a
functional of densities and collective velocities instead of
dealing with the impracticable Eq. (\ref{eq3}).
To implement this scheme, it is  quite clear that a sound
approximation is needed
for the imaginary-time action. This means a realistic
energy barrier and a simple yet reliable choice of the integration path
$q(\tau)$.

Let us now work out in some detail the  case in which
only one collective coordinate $\delta$ is considered.
It is then a simple task to  minimize the imaginary-time effective
action:
\begin{equation}
S(T)= \oint d \tau {\cal L}[\delta(\tau)]=
\int_{-\tau_p/2}^{\tau_p/2} d \tau \left[\frac{1}{2}M(\delta)
\dot{\delta}^2+\Delta\Omega(\delta)\right] \; .
\label{eq5}
\end{equation}
As indicated, the effect of continuing the action to imaginary time  is
to invert the `potential', i.e., $\Delta\Omega \rightarrow
-\Delta\Omega$ in the
equation of motion, and the identification of $k_B T$ with $\hbar/
\tau_p$; the path $\delta(\tau)$
defined in imaginary time $\tau$ has to fulfill the periodic boundary
condition $\delta(-\tau_p/2)=\delta(\tau_p/2)$. This is illustrated in
the schematic Figure \ref{fig1}.
We have supposed that the collective mass $M$ depends on $\delta$, which
will be the practical case. Imposing the extremum
condition on the action yields the following equation of motion for
$\delta(\tau)$:
\begin{equation}
M(\delta)\ddot{\delta} + \frac{1}{2} \frac{d M}{d
\delta} \dot{\delta}^2=\frac{d \Delta\Omega}{d \delta} \; .
\label{eq6}
\end{equation}
Multiplying Eq. (\ref{eq6}) by $\dot{\delta}$ we have
\begin{equation}
\frac{d}{d\tau}\left[\frac{1}{2} M(\delta)\dot{\delta}^2 -
\Delta\Omega(\delta) \right]\,=\,0 \; .
\label{eq7}
\end{equation}
Thus
\begin{equation}
\frac{1}{2} M(\delta)\dot{\delta}^2 - \Delta\Omega(\delta) =
constant  \equiv -E
\label{eq8}
\end{equation}
with $\Delta \Omega_{max} \geq E \geq 0$.

Eq. (\ref{eq8}) has the trivial  solution
$\delta =\delta_0$ corresponding to the minimum of
$-\Delta \Omega$: the system is `at rest' at the bottom of the inverted
barrier potential well. In this case, $E = \Delta \Omega_{max}$, and the
integration of Eq. (\ref{eq5}) yields
\begin{equation}
S^Q = \frac{1}{\hbar} S_{min}(T) = \frac{\Delta\Omega_{max}}{\hbar}
\int_{-\tau_p/2}^{\tau_p/2} d \tau = \frac{\Delta\Omega_{max}}{k_B T}
\; .
\label{eq9}
\end{equation}
Thus, the trivial solution yields
the exponent for classical thermal activation, see Eq. (\ref{eq1}).
It means that within FIA, the transition between thermal and quantal
regime is smooth. For  $E < \Delta \Omega_{max}$, one has
to seek periodic solutions $\delta(\tau)$ whose
turning points $\delta_1$ and $\delta_2$ are such
that $\Delta\Omega(\delta_1) = \Delta\Omega(\delta_2) = E$
(see Fig. \ref{fig1}). Integrating
Eq. (\ref{eq8}) we get the period $\tau_p$:
\begin{equation}
\tau_p(E) = 2\int_{\delta_1(E)}^{\delta_2(E)}  d\delta
\sqrt{\frac{M(\delta)}{2[\Delta\Omega(\delta)-E]}}    \; .
\label{eq10}
\end{equation}
Using
\begin{equation}
d\tau =
\sqrt{\frac{M(\delta)}{2[\Delta\Omega(\delta)-E]}}\; d\delta
\label{eq11}
\end{equation}
the action Eq. (\ref{eq5})  becomes
\begin{equation}
S^Q(E) = \frac{2}{\hbar} \int_{\delta_1(E)}^{\delta_2(E)}  d\delta\,
[2\Delta\Omega(\delta)-E]
\sqrt{\frac{M(\delta)}{2[\Delta\Omega(\delta)-E]}}  \; .
\label{eq12}
\end{equation}
At $T = 0$,  $\tau_p = \infty$ and $E = 0$. In this case, the
solution to Eq. (\ref{eq8}) is the usual instanton \cite{Col77}, and
$S^Q(E=0)$ coincides with  the
 WKB approximation at zero energy \cite{Gal90}:
\begin{equation}
S^{WKB} \equiv \frac{2}{\hbar} \int_{\delta_1(E=0)}^{\delta_2(E=0)}
d\delta \, \sqrt{2 M(\delta)\Delta\Omega(\delta)} \; .
\label{eq13}
\end{equation}

The crossover temperature is obtained by equating
Eqs. (\ref{eq1}) and (\ref{eq2}) we get:
\begin{equation}
S^Q(E=\Delta\Omega_{max}) = \frac{\Delta\Omega_{max}}{k_B T^*} \; .
\label{eq14}
\end{equation}
Taking a trajectory $\delta$ corresponding to
$E\approx\Delta\Omega_{max}$ and using Eq. (\ref{eq12}),
 we can write
\begin{equation}
S^Q(E \approx \Delta\Omega_{max}) \approx \frac{2}{\hbar}
\Delta\Omega_{max}
\int_{\delta_1(E \approx \Delta\Omega_{max})}^{\delta_2(E \approx
\Delta\Omega_{max})}
 d \delta\;
\sqrt{\frac{M(\delta)}{2\left[\Delta\Omega(\delta)-E\right]}} \; .
\label{eq15}
\end{equation}
The comparison to Eq. (\ref{eq10})  yields
 $\hbar/(k_B T^*) = \tau_p(E \approx \Delta\Omega_{max})$.
An analytical expression for $T^*$ is obtained by expanding
Eq. (\ref{eq10}) around the maximum of $\Delta\Omega$
located at $\delta_0$. It reads
\begin{equation}
k_B T^* = \frac{\hbar}{2\pi}\sqrt{\left.-\frac{1}{M(\delta_0)} \frac{d^2
\Delta\Omega}{d \delta^2} \right|_{\delta_0}} \; .
\label{eq16}
\end{equation}
Eq. (\ref{eq16}) shows that  the value of $T^*$
is determined by {\em  small
variations around $\delta_0$}. This is a
well-known result that was suggested long ago by Goldanskii \cite{Gol59},
the validity of which goes beyond the simple model we have used to derive
it. Generally speaking, the crossover temperature is determined from the
frequency of the small amplitude oscillations around the minimum of the
inverted barrier potential well. When more realistic methods are
employed to generate the critical nucleation configurations,
like for example density functional theory
\cite{Xio89,Oxt92}, the problem becomes an infinite-dimensional one,
whose solution is quite a formidable task. Crossover temperatures for
thermally assisted quantum cavitation in liquid helium have been
determined only recently \cite{Gui96,Jez97b}.

Finally, we mention that $S^{WKB}$ can be used to estimate $T^*$ through
the expression \cite{Xio89,Bur93}
\begin{equation}
k_B T^*_{WKB} = \frac{\Delta\Omega_{max}}{S^{WKB}}  \; .
\label{eq17}
\end{equation}
\section{Energy barriers and imaginary-time actions}

As indicated at the introduction, we have considered two different
nucleation seeds, depending on whether there is some vorticity in the
experimental sample or not. The first one is a pure $^3$He droplet
embedded into the $^3$He-$^4$He mixture, and the second one is a
$^4$He vortex line filled with pure $^3$He.
In the following we work out in detail the energy barriers and imaginary
time actions making use of the
capillarity approximation in the case of a droplet, and of the hollow
core approximation in the case of a  vortex line. These are sound
approximations provided  nucleation occurs  near the two phase
equilibrium curve. Indeed, it
is well known that near this curve, the critical nucleation
configuration,
i.e. that corresponding to the maximum of the energy barrier, is large.
This makes negligible `finite size effects' such as interface
diffuseness and curvature corrections. This has been checked by comparing
the capillarity model results with density functional
calculations, which selfconsistently incorporate finite size effects
\cite{Gui95,Jez95,Jez97}. We recall that
experimentally, the process occurs near the saturation line;
otherwise, $\Delta x_{cr}$ would be large.
This is at variance with the process of
cavitation in liquid helium, which takes place near the
spinodal line\cite{Pet94,Lam98,Cau98},
where the cavitation seeds are small, pachydermic bubbles
and the capillarity approximation fails \cite{Gui96,Xio89,Mar95}.
In the case of nucleation triggered by $^3$He droplets, we will also
present the results obtained using a  density functional FIA as
described in Ref. \onlinecite{Jez97b}.

\subsection{$^3$He droplets}

Within the capillarity approximation,
the energy barrier of a $^3$He droplet of radius $R$ immersed in the
mixture is written as
\begin{equation}
\Delta \Omega(R) = 4 \pi R^2 \sigma  - \frac {4 \pi}{3} R^3 \rho_{30}
\,\Delta \mu_3 \; ,
\label{eq18}
\end{equation}
where $\sigma$ is the surface tension of the $^3$He-$^4$He interface,
 $\rho_{30}$ is the density of pure $^3$He inside the
droplet, and $\Delta \mu_3$ is the difference between the chemical
potential of $^3$He in the metastable mixture and in pure $^3$He. This
difference is negative when the mixture is stable and positive otherwise.
Upon minimization with respect to $R$, the maximum of
the energy barrier and the radius of the critical droplet $R_c$
are given by
\begin{eqnarray}
\Delta \Omega_{max} &=& \frac{4 \pi}{3} \sigma R_c^2 =
 \frac {16 \pi}{3} \frac{ \sigma^3}{(\rho_{30}\, \Delta \mu_3)^2}
\nonumber \\
R_c &=& 2 \sigma/(\rho_{30}\, \Delta \mu_3) \; .
\label{eq19}
\end{eqnarray}
Values of $R_c$ and $\Delta \Omega_{max}$ can be obtained using the
experimental values\cite{Sel69,Sat97,Ker62}
of the physical quantities entering their definition. In particular, the
experimental data\cite{Sel69} can be employed to write the above
equations as a function of $\Delta x$  using the approximation
$\Delta \mu_3 \approx (\partial \Delta \mu_3 / \partial x) \Delta x$
along the demixing line. For instance, at zero pressure $\Delta \mu_3
\sim 2.49\, \Delta x$ K. In view of the weak $T$ dependence of
these quantities in the range of temperatures  of experimental interest
(below 150 mK), it is legitimate to consider their values at
zero K. If nucleation proceeds thermally, the degree of
critical supersaturation at  fixed $P$ and $T$ is obtained by
solving the equation $1 = (V \cdot t)_{exp} J_T$. It determines
$\Delta x_{cr}$ from $\Delta \Omega_{max}$ (Eq. (\ref{eq19})):
\begin{equation}
\Delta \Omega_{max} = k_B T \ln [(V \cdot t)_{exp} J_{0T}] \; .
\label{eq20}
\end{equation}
We have used  as prefactor an attempting frequency per unit volume
$J_{0T} = \nu_0 / V_0 \sim (k_B T / h)/ (4 \pi
R_0^3/3)$. As pointed out, the prefactor depends on the
physics of the nucleation process, but the results are not very
sensitive to the precise value of the product $(V \cdot t)_{exp} J_{0T}$
(see the discussion after Eq. (\ref{eq30})). Yet, using the estimates
made at the end of Ref. \onlinecite{Jez97b} we have checked that in the
present case, where phase separation proceeds by the diffusion of $^3$He
atoms, the diffusion time to make a critical $^3$He droplet of $R_0 \sim
10 {\rm \AA}$ is, in the worst case, of the same order of magnitude than
the 'thermal period' $\tau_T = 2 \pi h/k_B T$. As thoroughly discussed by
Burmistrov et al\cite{Bur93}, care should be exerted when describing
later stages in the dynamical evolution of the critical droplet after
it reaches a mesoscopic size.
Taking for the experimental volume and time
$(V \cdot t)_{exp}$ those of Ref. \onlinecite{Lam98} and for $R_0 \sim
10 {\rm \AA}$, we obtain $\Delta x_{cr} \sim 13 \%$ at $T$ = 100 mK,
and $\sim 19 \%$ at $T$ = 50 mK. This result is an order of
magnitude larger than the experimental data.
This is a  drawback of using $^3$He droplets as
nucleations seeds detected inmediately after the first systematic
experiments\cite{Mik91,Sat92}.

Let us now calculate the crossover temperature.
In the following, we write $R = R_c + \delta$,
where here $\delta$ measures the displacement of the surface of the drop from
the value corresponding to the critical configuration. In this way, the
dynamics of the nucleation process is described by the time
evolution of $\delta$. The energy barrier then reads
\begin{equation}
\Delta \Omega(\delta) = 4 \pi (R_c + \delta)^2 \sigma - \frac {4 \pi}{3}
(R_c + \delta)^3 \rho_{30} \,\Delta \mu_3 \; .
\label{eq21}
\end{equation}
The next step is to express the imaginary-time Lagrangian in terms of the
displacement $\delta$ and its time derivative $\dot{\delta}$. Since the
`potential part' is just $- \Delta \Omega(\delta)$, we have only to
derive the kinetic energy term. This last is given by
\begin{eqnarray}
E_{kin} & = & \frac{m_3}{2} \int d \vec{r}\, \rho_3(\vec{r}, t)\,
 \vec{u}_3^{\,2}(\vec{r}, t)
+ \frac{m_4}{2} \int d \vec{r} \,\rho_4 (\vec{r}, t)
 \vec{u}_4^{\,2}(\vec{r}, t)   \nonumber \\
& \equiv & \frac{1}{2} M(\delta)\, \dot{\delta}^2 \; ,
\label{eq22}
\end{eqnarray}
where $ \rho_q(\vec{r}, t)$ and $\vec{u}_q(\vec{r}, t)$ are
respectively, the particle density and  the collective velocity
of isotope $q$.
The collective mass  $M(\delta)$ is determined as follows. For a
given configuration characterized by a $\delta$ value, the densities of
the system are written as:
\begin{eqnarray}
\rho_4(r) & \equiv & \rho_{4\delta}(r) =
\rho_{4m}  [ 1 - \Theta (R_c + \delta - r)  ]
                 =  \rho_{4c} (r - \delta )  \nonumber \\
\rho_3(r) & \equiv & \rho_{3\delta}(r) =
\rho_{3m}  [ 1 - \Theta (R_c + \delta - r)  ]
+ \rho_{30}  \Theta (R_c + \delta - r)
                 =  \rho_{3c} (r - \delta )  \, ,
\label{eq23}
\end{eqnarray}
where $\rho_{qm}$ with $q$ = 3,4 is the particle density of each
isotope in the metastable phase ($x = \rho_{3m}/(\rho_{3m}+\rho_{4m})$),
and the subscript $c$ refers to the critical configuration. This is
schematically illustrated in Fig. \ref{fig2}. The
dynamics comes into Eqs. (\ref{eq23}) through the time dependence of
$\delta$, i.e., $\rho_q(r,t) = \rho_{qc}(r -\delta(t))$. Thus,
\begin{equation}
\rho_q (\dot{r},t) = - \dot{\delta} \, \frac{d}{dr } \rho_{qc}(r) \; .
\label{eq24}
\end{equation}
Assuming spherical symmetry during the process of growing,
the densities and
collective velocities only depend on the modulus of $\vec{r}$, and
 it is possible to write $u_q(r,t)$ as a function of $\delta$ and
$\dot{\delta}$ formally integrating the continuity equation \cite{Bli88}
\begin{equation}
\frac{\partial \rho_q}{\partial t} + \nabla ( \rho_q \vec{u}_q) = 0 \; .
\label{eq25}
\end{equation}
This yields
\begin{equation}
u_q(r,t) = - \frac{1}{r^2 \rho_q(r,t)}
\int^r_0 s^2 \dot{\rho_q}(s,t) d s \;  .
\label{eq26}
\end{equation}
Using Eqs. (\ref{eq23}) and (\ref{eq26}) the mass parameter becomes
\begin{eqnarray}
M(\delta) &=&  4 \pi  \left[m_4 \rho_{4m} +
 m_3 \frac{(\rho_{30}-\rho_{3m})^2}{\rho_{3m}}\right]
 (R_c + \delta)^3 \nonumber \\
& \equiv & 4 \pi {\cal M}  (R_c + \delta)^3  \; .
\label{eq27}
\end{eqnarray}
The same expression has been derived in Refs.
\onlinecite{Bur96,Bur93} using a different method.

By inserting the expressions for $\Delta\Omega(\delta)$ and $M(\delta)$
in Eq. (\ref{eq16}),
we straightforwardly obtain the crossover temperature
\begin{equation}
k_B T^* = \frac{\hbar}{2\pi}\sqrt{\frac{1}{{\cal M}}
\,\frac{ 2 \sigma}{R_c^3}} \; .
\label{eq28}
\end{equation}
Furthermore, Eq. (\ref{eq17}) yields
\begin{equation}
T^*_{WKB} = \sqrt{\frac{2}{3}}\, \frac{512}{405}\, T^* \approx \,1.03
\,T^* \; .
\label{eq29}
\end{equation}
>From the $\Delta x_{cr}$ values quoted below Eq. (\ref{eq20}),
 we estimate $T^* \sim$ 16 mK for  $\Delta x_{cr}$ = 13 \%, and
$T^* \sim$ 28 mK for  $\Delta x_{cr}$ = 19 \%. These crossover
temperatures are compatible with those experimentally
found\cite{Sat92}.
We want to stress that these estimates are quantitative provided $\Delta
x$ is small, i.e., far enough from the spinodal line in the $P-x$
plane, which  is determined by the condition\cite{Gui95} $(\partial
\mu_3 / \partial x)_{P,T} = 0$.  Indeed, as the nucleation barrier
approaches
to zero at the spinodal line, so does $T^*$, and the law $T^* \sim
(\Delta x)^{3/2}$ following from Eqs. (\ref{eq19}) and (\ref{eq28})
eventually breaks down. A similar breakdown occurs in the cavitation
process\cite{Xio89,Gui96b}. Fig. \ref{fig3} displays $T^*(\Delta x)$
obtained from equation (\ref{eq28}) at $P=0$ and 1 bar (dashed lines).
The values of
$\rho_{3m}$ and $\rho_{4m}$ entering Eq. (\ref{eq27}) when
$x$ is much larger than $x_s$ have been calculated by means of
 the density functional of
Ref. \onlinecite{Bar97}. Had we used the values at the saturation line,
the resulting $T^*$ would have been a factor of two smaller for $\Delta
x$ larger than $\sim 5\%$.

We have also obtained  $T^*(\Delta x)$ using density
functional theory to describe the critical nucleation clusters, and the
FIA for the nucleation dynamics \cite{Jez97b}. 
The results are displayed in Fig. \ref{fig3} (solid lines).
For completeness, and to
clarify some aspects underlying our way of obtaining the collective mass
in the capillarity approximation,
we present in the Appendix the equations of motion used to determine 
$T^*$.

It can be concluded from Fig. \ref{fig3} that the $T^*(\Delta x)$ function
obtained using a density functional that accurately describes the
thermodynamical properties of the liquid mixture \cite{Bar97}
has a maximum of $\sim$ 25 mK at $\Delta x$
$\sim$ 18 \%. Consequently, at $P=0$ above $T \sim $ 25 mK the
nucleation of the $^3$He-rich phase always proceeds thermally.
At variance with the capillarity method, the density functional
approach is able to realistically describe critical nucleation
configurations from the saturation to the spinodal point. It is
worth to notice that the crossover temperatures obtained using the
density functional plus FIA tend to zero at the saturation and spinodal
points, whereas those obtained from the capillarity approximation
vanish only at saturation.
The calculated spinodal $\Delta x$ value at $P \sim 0$ is $\sim 25
\%$ (see Ref. \onlinecite{Gui95} and also Fig. \ref{fig3}).
The crudeness of the capillarity approximation when applied too far from
the coexistence line can be  assessed comparing the density profiles
of the critical nucleation configuration of the density
functional method versus the capillarity approximation. Fig. \ref{fig4}
shows them at $P = 0$ for $\Delta x_{cr}$ = 16 \%.

Figure \ref{fig3} simply tells us that if the mixture is brought to a
supersaturation degree $\Delta x$, nucleation will proceed by quantum
tunneling below $T^*(\Delta x)$. To determine
which of the $T^*(\Delta x)$  would correspond to a given experimental
situation, one has to determine the critical degree of
supersaturation $\Delta x_{cr}$, which corresponds to the largest one
the mixture can sustain before $^3$He droplets nucleate at an
appreciable rate. This is obtained by solving the equation
\begin{equation}
1 = (V \cdot t)_{exp} J
\label{eq30}
\end{equation}
taking either $J=J_T$ for $T \geq T^*$ or $J=J_Q$ for $T \leq T^*$, and
using as a prefactor $J_{0Q}$ the estimate $k_B T^*/(h V_0)$, where
$V_0 = 4 \pi R^3_0/3$ with $R_0 = 10 {\rm \AA}$. Within the more
accurate density functional FIA and using as experimental
$(V \cdot t)_{exp}$ two extreme values, namely $10^{14} {\rm \AA}^3$ sec
and $10^4 {\rm \AA}^3$ sec, at $P = 0$ one gets $\Delta x_{cr}
\sim 18 \%$ and $T^* \sim$ 25 mK, and  $\Delta x_{cr}
\sim 16 \%$ and $T^* \sim$ 23 mK respectively, and very similar values
at $P = 1$ bar. The maximum of the thermal nucleation barriers
$\Delta\Omega_{max}$ are $\sim 1 K$ and $\sim 0.5$ K, respectively.

\subsection{$^4$He vortex lines filled with $^3$He}

The possibility of considering vortex lines as seeds of a kind of
heterogeneous nucleation in the mixture stems from the fact that the
experimental sample, if treated in conventional fashion, is likely
permeated {\em ab initio} by quantized vortices stabilized by surface
pinning \cite{Aws84}. These vortex lines, whose core is filled with
$^3$He, are stable for $x \leq x_s$, and metastable otherwise
\cite{Jez95,Jez98}. Within the hollow core model (HCM), is it very simple
to obtain their structure. The energy per unit length of vortex
line is
\begin{equation}
\Omega(R) = 2 \pi R \,\sigma  -  \pi R^2 \rho_{30} \,\Delta \mu_3
+ \pi n^2\ \frac{\hbar^2}{m_4} \rho_{4m} \ln \left(\frac{R_{\infty}}
{R}\right) \; .
\label{eq31}
\end{equation}
In this equation, $R$ is the radius of the vortex core, $R_{\infty}$ is
a large enough radius at which velocity vanishes, $n = 1, 2, \ldots$ is
the quantum circulation number, and the remaining
variables have the same meaning as before. Minimization with
respect to $R$ yields at most  two extrema corresponding to the
values
\begin{equation}
R = 2 R_0 \frac{\mu_c}{\Delta \mu_3} \left[ 1 \pm
\sqrt{1 - \frac{\Delta \mu_3}{\mu_c}}\right]  \; ,
\label{eq32}
\end{equation}
where
\begin{eqnarray}
R_0 &\equiv& \frac{n^2 \hbar^2 \rho_{4m}}{2 \sigma m_4} \\ \nonumber
\mu_c &\equiv& \frac{\sigma^2 m_4}{2 n^2 \hbar^2 \rho_{30}\, \rho_{4m}}
\; .
\label{eq33}
\end{eqnarray}
$ R_0 = \lim_{\Delta \mu_3 \rightarrow 0} R $
is the radius of the vortex
core at the demixing line, i.e., $\Delta \mu_3 = 0$. If the vortex is
in the stable region of the $P - x$ plane, $\Delta \mu_3$ is negative,
only the minus sign in Eq. (\ref{eq32}) makes sense and the
corresponding $R$ defines the radius of the stable vortex. If
$\Delta \mu_3 > 0$,  Eq. (\ref{eq32}) yields two core radii. The
smallest
one $R_<$ corresponds to a minimum of $\Omega(R)$, and the largest one
$R_>$ to a maximum. This is, of course, the situation for given
$(P, T, \Delta x)$ values. Consequently, {\em ab initio} stable vortex
lines become metastable as the mixture is driven into the supersaturated
region. These metastable vortex lines can trigger phase separation by
thermal fluctuations if an energy per unit length equal to
$\Delta\Omega_{max} = [\Omega(R_>)-\Omega(R_<)]$ is subministrated to
them. As before, calling $R_c$ the radius $R_>$ of the critical
configuration and defining $R \equiv R_c + \delta$, the energy barrier
per unit length reads
\begin{equation}
\Delta\Omega(\delta) = 2 \pi (R_c + \delta)\sigma
-  \pi (R_c + \delta)^2 \rho_{30}\, \Delta \mu_3
+ \pi n^2\ \frac{\hbar^2}{m_4} \rho_{4m} \ln \left(\frac{R_{\infty}}
{R_c + \delta}\right) - \Omega(R_<) \; ,
\label{eq34}
\end{equation}
where $\Omega(R_<)$ is a constant for fixed $(P, T, \Delta x)$.

It is interesting to notice from Eq. (\ref{eq32}) that if $\Delta \mu_3
> \mu_c$
the vortex line loses its metastable character. Indeed, when
$\mu_3 = \mu_c$ one has $R_> = R_<$: the energy barrier disappears and
the vortex core freely expands triggering  phase separation.
At $P = 0$, taking
$n=1$, $\rho_{4m} \sim 0.020 {\rm \AA}^{-3}$ and the experimental values
for the other variables, one gets $R_0 \sim 7.6 {\rm \AA}$ and
$\mu_c \sim 0.032$ K which yields $\Delta x_{cr} \sim 1.3 \%$.
It is remarkable that this  good agreement with experimental supersaturation
values  comes
from an extremely simple model. The smallness of the obtained
$\Delta x_{cr}$  fully justifies the use of the hollow core model
to calculate $\Delta x_{cr}$ and $T^*$.

To determine $T^*$ we proceed as in the preceeding subsection. Equation
(\ref{eq22}) gives now the kinetic energy per unit length, and the
expressions (\ref{eq23}) hold with $r$ being the cylindrical
radial coordinate. Again, the continuity equation in cylindrical
coordinates can be formally integrated yielding
\begin{equation}
u_q(r,t) = - \frac{1}{r \rho_q(r,t)}
\int^r_0 s \dot{\rho_q}(s,t) d s \;  .
\label{eq35}
\end{equation}
The collective mass for this geometry is readily obtained:
\begin{equation}
M(\delta) =
2 \pi {\cal M}  (R_c + \delta)^2 \ln
\left(\frac{R_{\infty}}{R_c +\delta}\right)  \; ,
\label{eq36}
\end{equation}
where ${\cal M}$ is defined in Eq. (\ref{eq27}).
In deriving the vibrational collective mass we have not taken
into account any contribution from the
circulation of the superfluid component, as only small
radial displacements have to be considered if one supposes that
phase separation is triggered by the expansion of the vortex core.
Finally, we have for the crossover temperature
\begin{equation}
k_B T^* = \frac{\hbar}{2\pi}\sqrt{\frac{1}{{\cal M}}
\,\frac{ \Delta \mu_3 \rho_{30} - R_0 \sigma/R_c^2}
{R_c^2 \ln \left(\frac{R_{\infty}}{R_c}\right)}}  \; .
\label{eq37}
\end{equation}
Taking $\ln(R_{\infty}/R_c) \sim 1$, one gets $T^* \sim 1$ mK for
$\Delta x \sim 1 \%$ at $P= 0$.

\section{Summary}

In this work we have calculated the crossover temperature and critical
supersaturation of $^3$He-$^4$He mixtures at very low temperatures
using a functional-integral method in conjuction with either a density
functional or a capillarity approximation description of  nucleation
configurations.
We have found that the existence of  vorticity in the mixture would
explain the small  $\Delta x_{cr}$ measured in experiments, but
fails to reproduce $T^*$, yielding values an order of
magnitude smaller than experiment.
Conversely, conventional nucleation seeds consisting in $^3$He droplets
do reproduce the experimental crossover temperatures but yield
$\Delta x_{cr}$ values an order of magnitude larger than experiment.

In spite of its failure, we feel that a detailed exposure of the merits
and drawbacks of the present method might help find alternative roads to
address
this longstanding problem and identify any missing physical ingredient
relevant to its solution. Besides, we think that our application of
the FIA to this problem is especially simple and transparent, which adds
a pedagogical value to it.

A line of possible improvement might consists in introducing from the
start a dynamics based on a zero rather than a first sound
hydrodynamic  description of
the harmonic vibrations leading to $T^*$. The density functional we have
used reproduces the Landau parameters of the mixture that are relevant
for such a zero sound description in the long wavelength limit;
among other quantities, $^3$He
effective mass and the incompressibilities of both isotopes have been
taken into account
as fitting quantities to fix the density functional parameters.
Moreover,  the nucleation barrier is known to be well described
by the new generation of static density functional  calculations
presented in Refs. \onlinecite{Gui96,Xio89,Jez97b}. It would be
quite natural  to introduce the zero sound dynamics much
along the fruitful procedure followed in nuclear
physics to describe nuclear collective excitations with
density-dependent  nucleon-nucleon efective interactions
\cite{Rin80}. Such a
program, which involves solving the Random-Phase-Approximation
equations in imaginary time, has  been formally developed to
study a problem having some similarities to ours, namely
spontaneous
nuclear fission below the fission barrier\cite{Rei81}. Here and there,
a practical implementation of this scheme is quite a formidable task,
especially in presence of vortices.

Finally, we would like to point out that the experience gathered
in the description of the nuclear giant monopole resonance indicates
that for this precise mode, first and zero sound dynamics yield
quite equivalent results, see chapter 13 of Ref. \onlinecite{Rin80}
and Ref. \onlinecite{Kri80}.
Thus, in the case of nucleation driven by $^3$He droplets,
as far as the problem of determining $T^*$ is amenable
to study spherically symmetric, radial vibrations of the system, we
expect that both dynamics yield rather similar $T^*$.

\section{Acknowledgments}
We would like to thank Eugene Chudnovsky for the many useful discussions
we have had on the general problem of quantum cavitation.
This work has been performed under grants from INFM (Italy),
PB95-1249 and PB97-1139 from DGICYT,
1998SGR00011 from  Generalitat de Catalunya, and by the IN2P3-CICYT
agreement.

\appendix
\section*{}

In this Appendix we give some hints about how we have derived  the 
dynamical
equations leading to $T^*$ in the case of $^3$He-$^4$He mixtures. 
We recall that by construction, the static density functional assumes
that $^3$He is in the normal phase, and all $^4$He is superfluid. 
For this reason, the
'two fluid' hydrodynamics of superfluids, as described for instance
in Ref. \onlinecite{Kha65}, is of no applicability here.  

The first step is to write the static density functional of the kind 
proposed by Dalfovo\cite{Dal89} (see also Refs 
\onlinecite{Gui95,Bar97}) in a Galilean invariant 
form\cite{Eng75,Gui95b}
 
\begin{eqnarray}
{\cal E}(\rho_q, \tau_q, \vec{j}_q) &=& 
\frac{1}{2 m_3} \tau_3 + \frac{1}{2 m_4} \tau_4 
+g_3(\rho_3, \rho_4) (\rho_3 \,\tau_3 - \vec{j}^{\,2}_3)
+g_4(\rho_3, \rho_4) (\rho_4 \,\tau_4 - \vec{j}^{\,2}_4)
\nonumber
\\
&+& g_{34}(\rho_3, \rho_4) (m^2_3\, \rho_3\, \tau_4 +
m^2_4 \,\rho_4 \,\tau_3 -2  m_3\, m_4 \,\vec{j}_3\cdot\vec{j}_4)
+ {\cal V}(\rho_3,\rho_4),
\label{eqa1}
\end{eqnarray}
where the kinetic energy $\tau_q$ and current $\vec{j}_q$ densities 
are defined in Ref. \onlinecite{Eng75}, and the functions $g_3$, $g_4$
and $g_{34}$ are not identically zero if the atoms have an effective mass
$m^*_q$ different from the bare mass
(we have taken $\hbar = 1$ thorough  the Appendix).
In the static case, the system is
invariant under time-reversal and the currents vanish\cite{Eng75}.
This allows to construct the $g$ functions from the static density 
functional without having to reparametrize it to reproduce the 
basic thermodynamical properties of the mixture. Yet, casting the
static density functional into a form like Eq. (\ref{eqa1}) leaves
plenty of freedom, as it is easy to convince oneself. 
The grand potential density is then built from the energy density, the
chemical potentials and particle densities: $\omega = {\cal E} - \sum_q
\mu_q \rho_q$.

One then applies to the $^3$He and $^4$He fluids a radial collective
velocity   $\vec{u}_q(\vec{r},t)$ ('boost')
deriving from velocity potential fields $s_q(\vec{r},t)$, 
$\vec{u}_q(\vec{r},t) = \nabla s_q(\vec{r},t)$ and takes into account
how the particle, kinetic energy and current densities are affected
by these boosts\cite{Eng75,Gui95b}. 
If $\vec{j}_3 = \vec{j}_4 = 0$ in the static system,
it is easy to check that the imaginary-time Lagrangian can be written as

\begin{equation}
{\cal L} =
 \sum_q \left [m_q \dot{\rho}_q s_q 
-\frac{1}{2} m_q \rho_q (\nabla s_q)^2\right ]
- \frac{1}{2} G_{34} (\nabla s_3 - \nabla s_4)^2
+ [\omega(\rho_q,\tau_q) -\omega(\rho_{qm},\tau_{qm})] \,\,\, ,
\label{eqa2}
\end{equation}
where $\frac{1}{2}\, G_{34} \equiv m_3^2 \,m_4^2 \,\rho_3 \,\rho_4 \,g_{34}$.
The imaginary-time
Hamiltonian $\cal{H}$ is readily obtained 
from $\cal{L}$\cite{Gui96,Jez97b}

\begin{equation}
{\cal H} = 
 \sum_q m_q \rho_q (\nabla s_q)^2 
+ \frac{1}{2} G_{34} (\nabla s_3 - \nabla s_4)^2
- [\omega(\rho_q,\tau_q) -\omega(\rho_{qm},\tau_{qm})] \,\,\, ,
\label{eqa3}
\end{equation}
and then the Hamilton equations 

\begin{eqnarray}
 m_q \dot{\rho}_q &=& \frac{\delta {\cal H}}{\delta s_q} 
\nonumber
\\
 m_q \dot{s}_q &=& - \frac{\delta {\cal H}}{\delta \rho_q} 
\label{eqa4}
\end{eqnarray}
yield, respectively, the continuity and
equations of motion:

\begin{eqnarray}
 m_3 \dot{\rho}_3 &=& -m_3 \nabla (\rho_3 \vec{u}_3) -
\nabla[G_{34} (\vec{u}_3 - \vec{u}_4)]
\nonumber
\\
 m_4 \dot{\rho}_4 &=& -m_4 \nabla (\rho_4 \vec{u}_4) +
\nabla[G_{34} (\vec{u}_3 - \vec{u}_4)]
\label{eqa5}
\end{eqnarray}

\begin{eqnarray}
 m_3 \frac{d\vec{u}_3}{dt} &=& 
- \nabla \left\{\frac{1}{2} m_3 \vec{u}_3^{\,2}
+\frac{\partial G_{34}}{\partial \rho_3} (\vec{u}_3 -\vec{u}_4)^2
-\frac{\delta \omega}{\delta \rho_3}\right\}
\nonumber
\\
 m_4 \frac{d\vec{u}_4}{dt} &=& 
- \nabla \left\{\frac{1}{2} m_4 \vec{u}_4^{\,2}
+\frac{\partial G_{34}}{\partial \rho_4} (\vec{u}_3 -\vec{u}_4)^2
-\frac{\delta \omega}{\delta \rho_4}\right\} \,.
\label{eqa6}
\end{eqnarray}
The above equations are quite similar to those obtained in Ref. 
\onlinecite{Kri80} to describe nuclear collective vibrations,
and at this point it is appropriate to recall the last paragraph
of Sect. IV. Linearizing Eqs. (\ref{eqa5},\ref{eqa6}) is 
straightforward and proceeds as outlined in Ref. \onlinecite{Jez97b}.
We do not write down the resulting equations, which are 
more involved than those given in the above reference because here we have
kept in the equations the terms arising from the possible existence
of a $g_{34}$ contribution. Rather, we
finish the Appendix with a short discussion
about possible choices for the $g$ functions.  

The more obvious and simple one is to take $g_4 = g_{34} \equiv 0$
and $g_3 \rho_3 \equiv 1/(2 m^*_3)$, where $m^*_3(\rho_3,\rho_4)$ is the
$^3$He effective mass in the mixture\cite{Gui95,Dal89}.
This leads to linearized equations to determine $T^*$ which are
extremely cumbersome but still manageable\cite{Jez97b}. They
have also been employed to obtain the density functional results  presented 
in this work, and 
consistent with this choice is to write the collective kinetic energy 
in the capillarity approximation as indicated by Eq. (\ref{eq22}).

The more general expression 
$g_3 \rho_3 + m^2_4 \rho_4 g_{34} \equiv 1/(2m^*_3)$ holds
if one uses the accurate parametrization\cite{Gui95,Dal89}
\begin{equation}
\frac{1}{2 m^*_3} = \frac{1}{2 m_3} 
\left [ 1 - \frac{\rho_3}{\rho_{3c}} - \frac{\rho_4}{\rho_{4c}} \right ]^2
\,\,\, ,
\label{eqa7}
\end{equation} 
where the fitting constants $\rho_{3c}$ $\rho_{4c}$ can be found in the 
above references.
It can be checked that  setting $g_4 = 0$ and splitting the above relation 
into 

\begin{eqnarray}
 g_3(\rho_3,\rho_4) &=& 
\frac{1}{2 m_3} \left [ \frac{\rho_3}{\rho^2_{3c}} - \frac{2}{\rho_{3c}}
+ \frac{ 2 \rho_4}{\rho_{3c}\rho_{4c}} \right]
\nonumber
\\
 g_{34}(\rho_3,\rho_4) &=& 
\frac{1}{2 m_3 m^2_4} \left [ \frac{\rho_4}{\rho^2_{4c}} - \frac{2}{\rho_{4c}}
\right] \,\,\, ,
\label{eqa8}
\end{eqnarray}
the good fit to the experimental $m^*_3$ is retained, and  
\begin{equation}
 \frac{m_4}{m^*_4} = 1 + \frac{m_3}{m_4} 
 \left [ \frac{\rho_4}{\rho^2_{4c}} 
+ \frac{ 2 \rho_3}{\rho_{3c}\rho_{4c}} - \frac{2}{\rho_{4c}}
\right] \rho_3 \,\,\, ,
\label{eqa9}
\end{equation}
which yields $m^*_4/m_4 = 1.31$  for one $^4$He impurity in normal 
liquid $^3$He  at saturation. This value is in agreement 
with the experimental one\cite{He98}.

\begin{figure}
\caption{Schematic barrier $\Delta\Omega(\delta)$ and inverted barrier
well.}
\label{fig1}
\end{figure}
\begin{figure}
\caption{Schematic figure to illustrate the calculation of the
collective mass.}
\label{fig2}
\end{figure}
\begin{figure}
\caption{$T^*$ (mK) as a function of $\Delta x (\%)$.
Dashed lines are the results obtained within the capillarity
approximation, and solid lines are the results of a density functional
plus FIA calculation.
The upper solid and dashed curves correspond to $P = 1$ bar, and the the
lower ones to $P = 0$.
}
\label{fig3}
\end{figure}
\begin{figure}
\caption{Density profile of the critical nucleation configuration
obtained using either a density functional method (thick lines), or the
capillarity approximation (thin lines). The values correspond to
$P = 0$ and $\Delta x = 16 \%$
}
\label{fig4}
\end{figure}
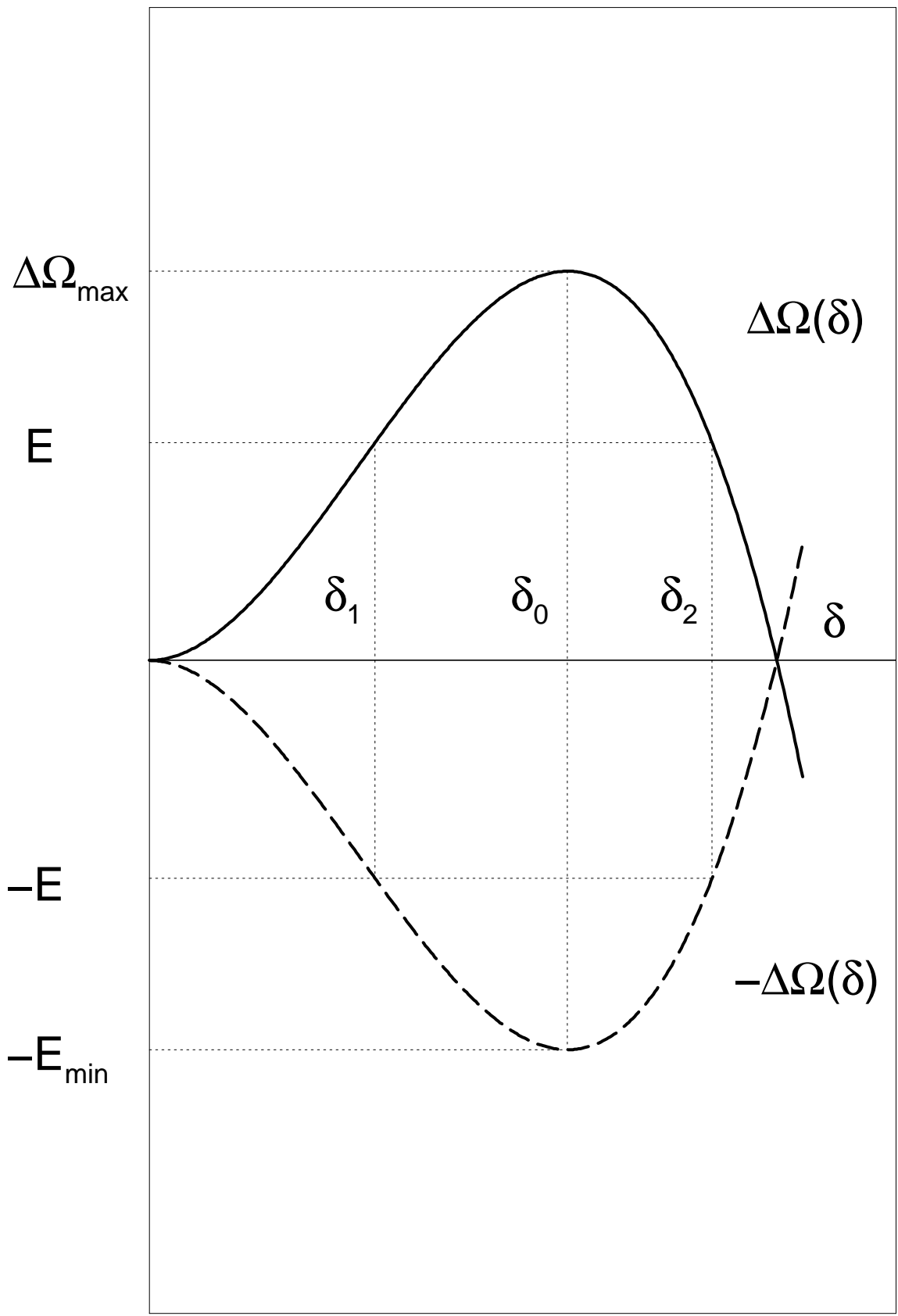
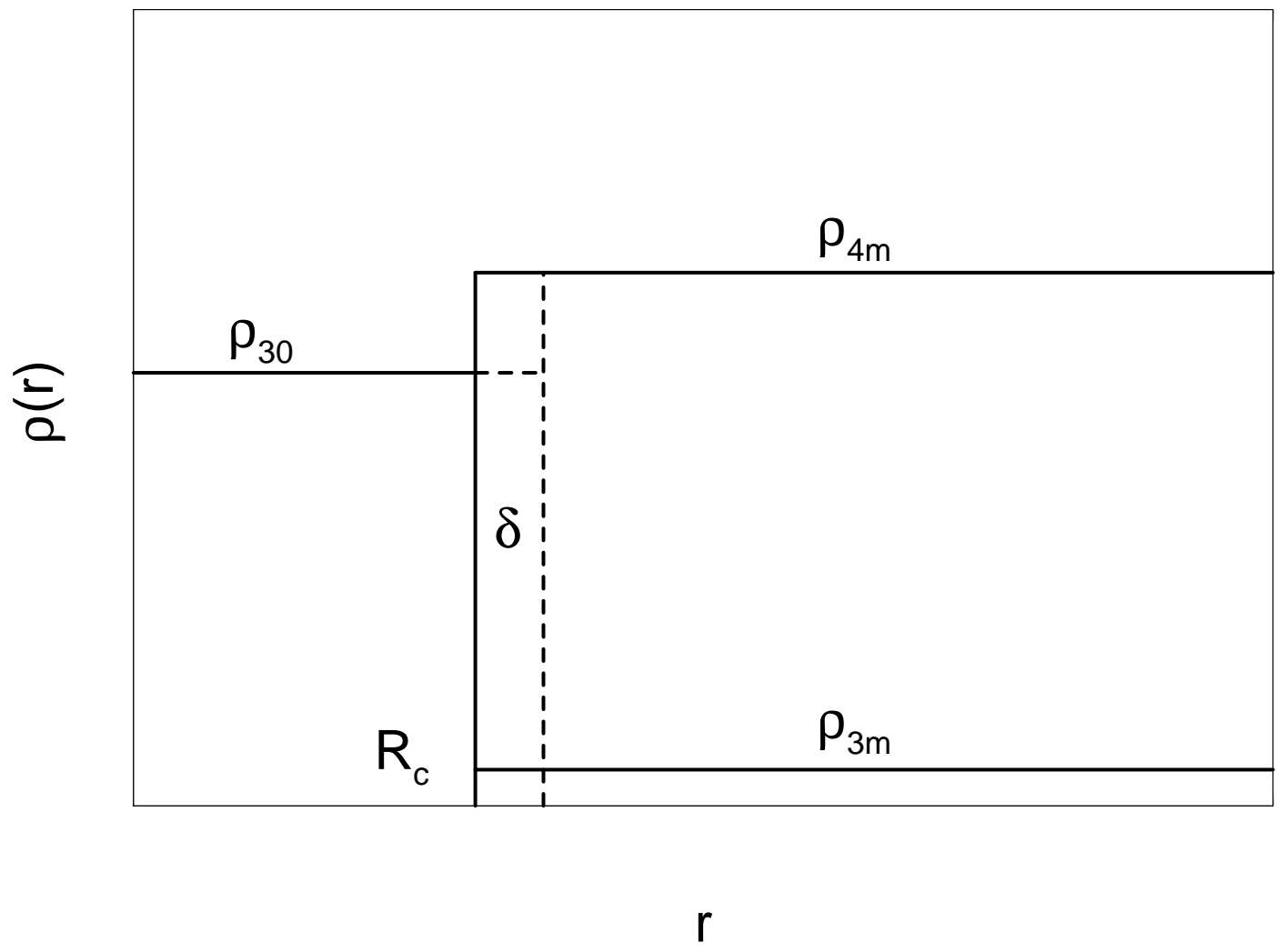
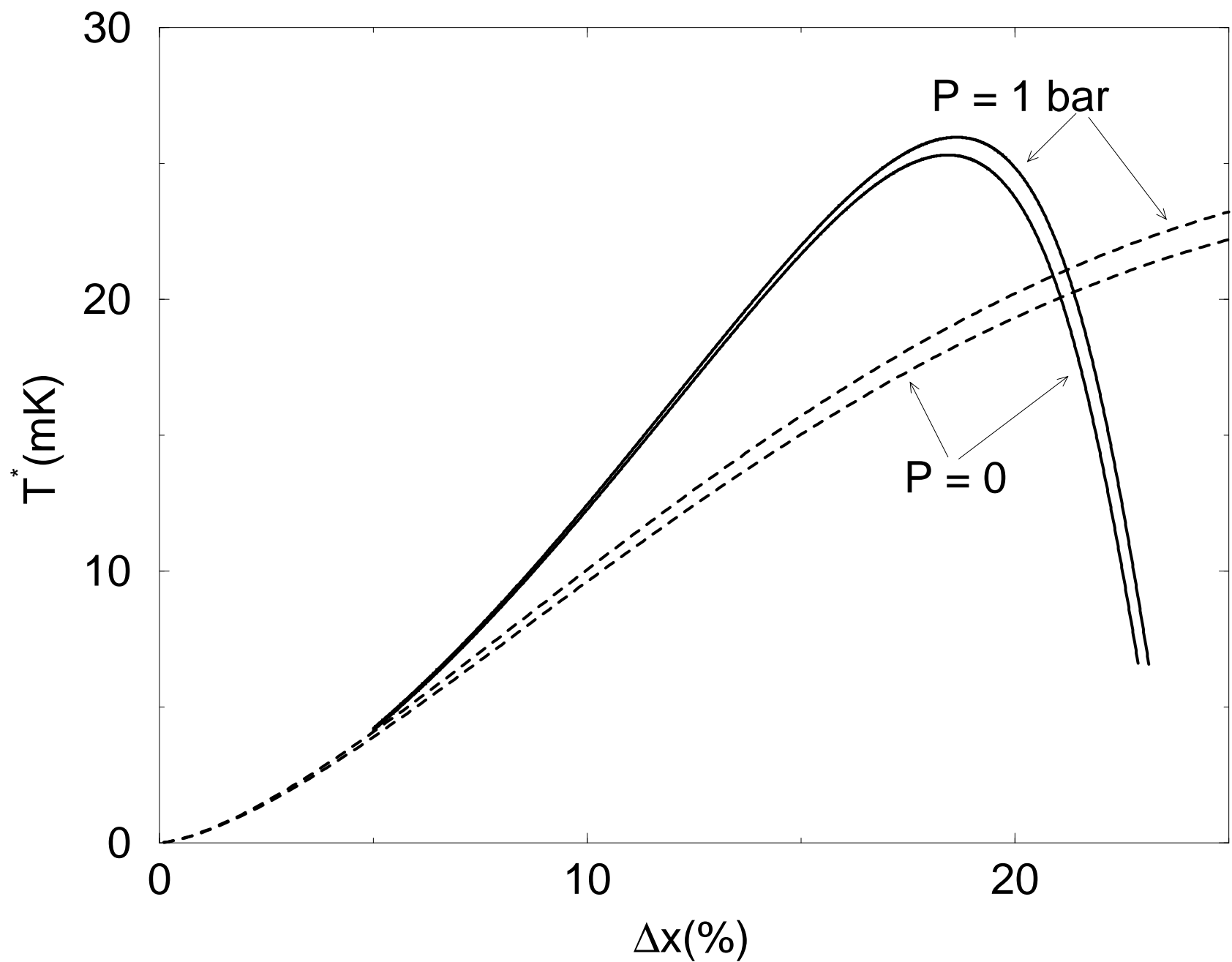
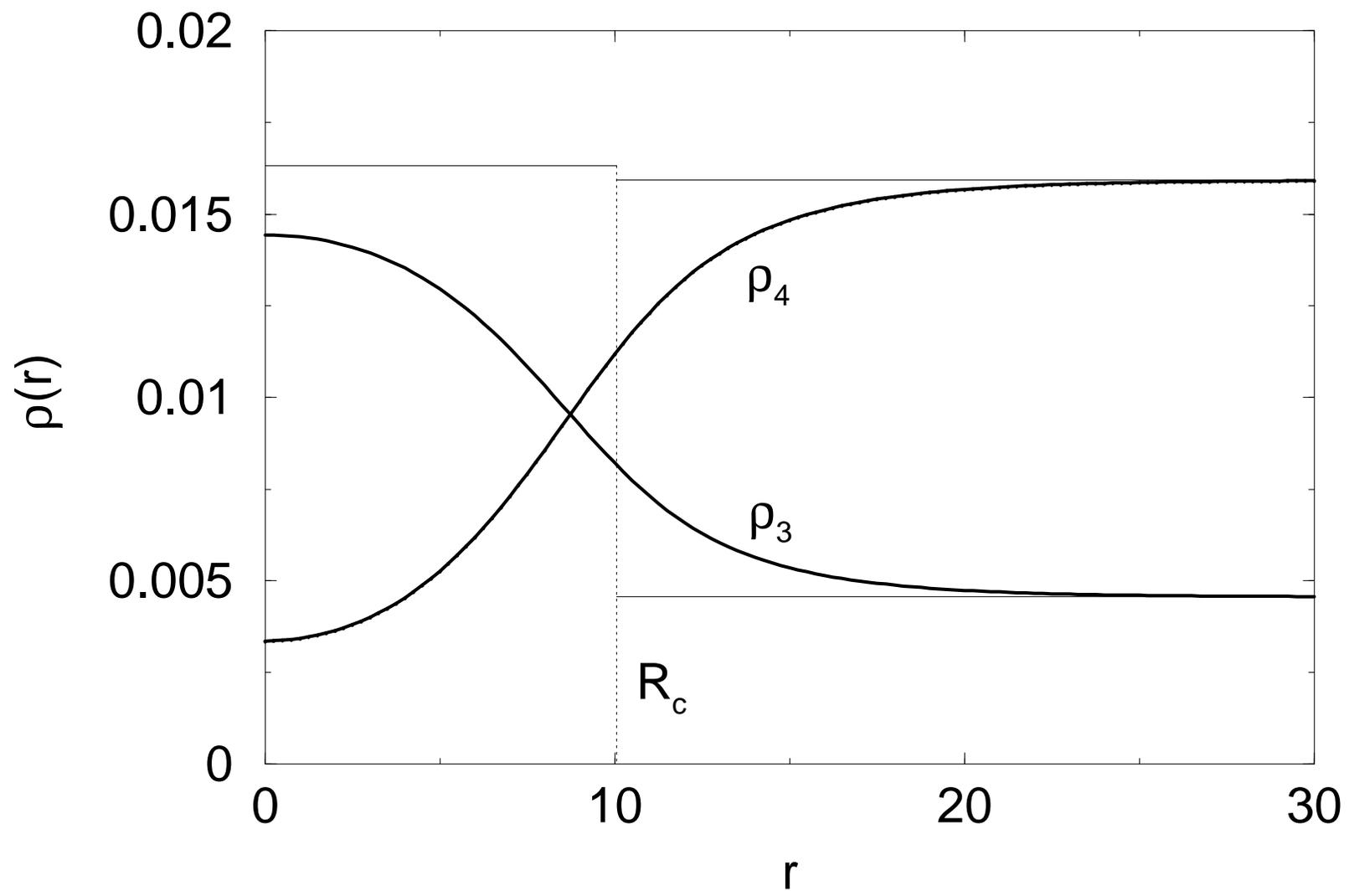
\end{document}